\documentclass[aps,prl,twocolumn,superscriptaddress,amsmath,amssymb,showpacs]{revtex4}
\usepackage{color}
\usepackage{graphics}
\usepackage{bm}
\begin{document}

\title{Selective coherent x-ray diffractive imaging of displacement fields in (Ga,Mn)As/GaAs periodical wires}

\author{A.~A.~Minkevich}
\email[]{andrey.minkevich@iss.fzk.de}
\author{E.~Fohtung}
\author{T.~Slobodskyy}
\author{M.~Riotte}
\author{D.~Grigoriev}

\affiliation{Institute for Synchrotron radiation,
Forschungszentrum Karlsruhe, 76344 Eggenstein-Leopoldshafen,
Germany}

\author{M.~Schmidbauer}

\affiliation{Leibniz-Institut f\"ur Kristallz\"uchtung, Max-Born-Stra{\ss}e 2,
D-12489 Berlin, Germany}

\author{A.~C.~Irvine}
\affiliation{Microelectronics Research Centre, Cavendish
Laboratory, University of Cambridge, Cambridge CB3 0HE, United
Kingdom}

\author{V. Nov\'{a}k}

\affiliation{Institute of Physics of the ASCR, Cukrovarnicka 10,
162 00 Praha, Czech Republic}

\author{V.~Hol\'{y}}

\affiliation{Department of Condensed Matter Physics, Faculty of
Mathematics and Physics, Charles University, Ke Karlovu 5, 121 16
Praha, Czech Republic}

\author{T.~Baumbach}

\affiliation{Institute for Synchrotron radiation,
Forschungszentrum Karlsruhe, 76344 Eggenstein-Leopoldshafen,
Germany}

\date{\today}

\begin{abstract}

\noindent Coherent x-ray diffractive imaging is extended to high
resolution strain analysis in crystalline nanostructured devices. The application potential is demonstrated by determining the
strain distribution in (Ga,Mn)As/GaAs nanowires. By separating diffraction signals in reciprocal spaces, individual parts of the device could be reconstructed independently by our inversion procedure. We demonstrate the method to be effective for material specific reconstruction of strain distribution in highly integrated devices.

\end{abstract}

\pacs{61.05.cp, 62.20.-x, 42.30.Rx, 75.50.Pp}% PACS, the Physics and Astronomy
                             % Classification Scheme.

\maketitle

%Introduction

Nanoscopic elastic strain analysis is crucial for understanding
physical properties and fabrication processes of crystalline
nanostructures and devices. In epitaxial nanostructures, strain distribution is strongly correlated to their
shape and composition. It is also an important driving force for self-assembling processes \cite{Schmidt07}.

In conventional x-ray diffraction experiments information about
shape, composition and strain is obtained \textit{indirectly} by
quantitative comparison of measured and simulated reciprocal-space
distributions of scattered intensity \cite{Holy99}. Simulations are
based on a model with  preliminary assumptions about detailed
shape, composition and strain profile, which will be confirmed by
the experimental results as long as the model structure fits to the reality.

A \emph{direct} determination of the density of the scattering centers from
the reciprocal-space distribution of the scattered intensity is
however not possible, unless a special \emph{phase-retrieval method} is
used. This phase-retrieval procedure known as coherent x-ray diffractive imaging (CXDI) has a potential to become a powerful technique for
the structural characterization of small objects and micro-sized
semiconductor devices \cite{Robinson09, Miao99, Chapman06, Minkevich07, Pfeifer06, Marchesini03, Williams06PRL, Thibault08}. The method principle is based on the iterative loop of direct and inverse Fourier transformation
(FT) (towards the experimental intensity distribution and back to the sample space) and it may refine the genuine object density distribution $\tilde{\varrho}({\bm r})$ even by starting from a model which is far from reality \cite{Fienup82, Elser03}.

In case of crystalline nano-objects the distribution of the scattered amplitude in
reciprocal space around a reciprocal lattice point (RLP) is given by
\begin{equation}
E({\bm Q}) \sim \int {\rm d}^3{\bm r} \varrho_{\bm h}({\bm r}) {\rm
e}^{-{\rm i}{\bm h}.{\bm u}({\bm r})}{\rm e}^{-{\rm i}({\bm
Q}-{\bm h}).{\bm r}}.
\end{equation}
Here, ${\bm h}$ and ${\bm Q}$ denote the reciprocal-lattice vector
and the position vector in reciprocal space (scattering
vector), respectively, $\varrho_{\bm h}({\bm r})$ is the
corresponding Fourier component of electron density and  ${\bm
u}({\bm r})$ is the displacement field with respect to the
non-strained lattice.

Once the phase of the scattered radiation is known, direct space
amplitude $\varrho_{\bm h}({\bm r})$ and phase ${\bm h}.{\bm
u}({\bm r})$ can be obtained by inverse FT, within the validity of the kinematical scattering theory
(first Born approximation).

An important step in the phase-retrieval loop is the application of
the direct(real)-space constraints (see~[\onlinecite{Fienup82}],
among others). Usually, the direct-space constraint has the form
of an \emph{a-priori} known support function determining the
region in direct space, where the complex electron density
$\tilde{\varrho}({\bm r})=\varrho_{\bm h}({\bm r}) \exp(-{\rm
i}{\bm h}.{\bm u}({\bm r}))$
may differ from zero. The first demonstration of the possibility of strain imaging by phase retrieval using support constraint on particular example of weakly strained Pb nanocrystal was recently illustrated in~[\onlinecite{Pfeifer06}].
Later it has been shown, however, that the
knowledge of the support constraint alone is not generally sufficient for
the phase retrieval of diffraction pattern from highly strained
crystals~\cite{Minkevich08}, but the problem may be overcome by
introducing additional constraints. Such a suitable
constraint in direct space is the use of a-priori knowledge to limit
the allowed values of strain in the iterative
process~\cite{Minkevich07, Minkevich08}.

\begin{figure}                %Fig.
\resizebox{6.5cm}{3 cm}{\includegraphics{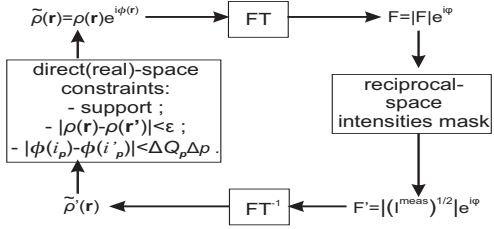}}
\caption{\label{fig_loop} Schematic diagram of the iterative loop of our phase retrieval algorithm. {\bf r} and {\bf r'} in "direct(real)-space constraints" box correspond to the neighbouring points within one constituent crystal and $\epsilon$ defines the threshold of electron density uniformity. $i_p$ and $i'_p$ are the neighbouring points along $p$-direction (normally lateral or vertical) and $\Delta p$ corresponds to the step along this direction (achieved resolution).}
\end{figure}

The diffraction amplitudes of compositional layered nanoobjects
can be understood as an interference sum of individual wave field
contributions scattered from the different material regions.
We may define firstly a non-perturbed state of the nanoobject by assuming for each material region idealized homogeneous composition and strain, spatially limited by the shapes of the individual nanostructures.
Secondly we define the perturbation by taking now the effects of non-uniform strain
relaxation and possible compositional fluctuations within previously homogeneous
layer region of the crystals into account.
Then, \textit{each} wave field contribution is the result of
convolution of two FTs: One is the individual reciprocal shape function
which leads to intensity decreases at least with $q^{-2}$ with
respect to the unperturbed RLP of that material.
The other one corresponds to the FT of $\tilde{\varrho}({\bm
r})$ containing the spatial fluctuations
in amplitude (compositional fluctuation) and phase (displacement
field). The spread of the later FT in frequency space is dependent
on the maximum gradient values of displacements in the layers and related to the strain-induced broadening of diffracted signal $\Delta{\bm Q}$ around RLP ($\Delta{\bm Q}_{p}$ - along particular $p$ direction). That fact allows to transform the "limited frequency constraint" into the direct space
constraint of limited component of the displacement gradient. Therefore, in addition to the standard support constraint we implement complementary direct-space constraints, namely, the uniformity of electron density (amplitudes) for each constituent crystal component and the requirement of
continuity of the displacement variations (of phase) (see Fig.~\ref{fig_loop}).

\begin{figure}                %Fig.
\resizebox{8.5cm}{6.5 cm}{\includegraphics{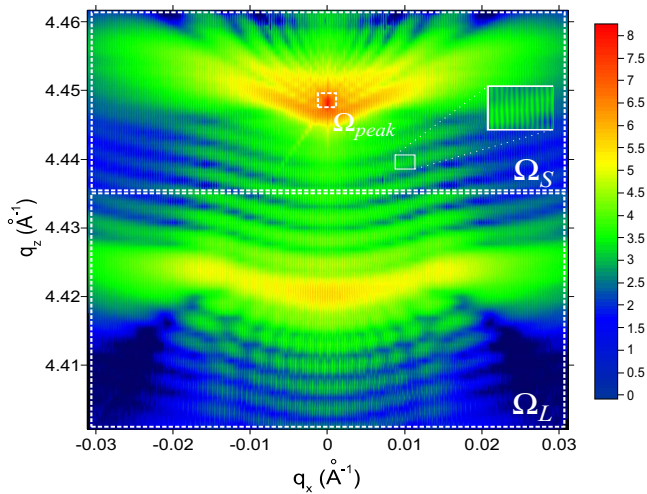}}%12
\caption{\label{figRSM004} (Color online) Measured reciprocal space map
(log 10 of intensities) from (Ga,Mn)As/GaAs periodical wires.
$\Omega_S$ and $\Omega_L$ confine the diffuse scattering regions
from GaAs and (Ga,Mn)As respectively.}
\end{figure}

Now, let us consider the constraint for the vertical frequencies. In
elastically strained-layer nano-objects the layer regions of
different composition have considerably different \emph{averaged}
vertical lattice parameters. The resulting mean vertical lattice
misfit between these regions leads to the separation of their
diffraction maximums in $q_z$-direction (see Fig.~\ref{figRSM004}) and to a
shift between their whole individual contributions in reciprocal
space by $\Delta{\bm h}$.

Mostly $\Delta{\bm h}$ is sufficiently large compared to the
maximum gradient of displacement field \emph{within} the
individual layers. If it is also sufficiently large compared to
the main contribution of the reciprocal shape functions,
reciprocal space regions are distinguishable, where the major
influence of scattered intensity stems mainly from one individual
material type. That consequently permits to reconstruct $\tilde{\varrho}$ for the individual constituent
parts of the nanoobject separately.

Experimentally, CXDI can be applied to individual nanoobjects like
wires and dots, if sufficient high photon flux density is
available. In order to gain in the radiation flux density, microfocusing
optics could be used~\cite{Schroer_PRL}. Another way to overcome the problem is the use of periodical objects. If the scattering coherent volume contains a
large number of equally spaced identical objects, we gain
constructive interference concentrating and amplifying diffracted
intensity into so-called grating rods \cite{Holy99}.
%More precisely, the coherent diffraction pattern is a product of the scattering pattern by a single object (envelope) and the so-called interference function represented by the lateral sequence of highly intense grating rods.
These rods are separated by $\Delta
q_x =2\pi/T$, where $T$ is a period size. The maximum intensity of
the scattered radiation in rods is given by
$I_{\rm exp}=I_1\frac{N}{N_c}N_c^2=I_1NN_c$
where $N$ is the number of the periods irradiated by the primary
wave, $N_c$ is the number of coherently irradiated periods, and
$I_1$ is the maximum intensity scattered from a single wire.
Therefore, if we use an ideally coherent primary wave, the
scattered intensity is substantially enhanced, since it is
proportional to $N^2$. Consequently sampling in reciprocal space
can be performed in a wider range, where measured intensities are
without coherent magnification already negligibly small.

\begin{figure}                %Fig.
\resizebox{8.5cm}{2.3 cm}{\includegraphics{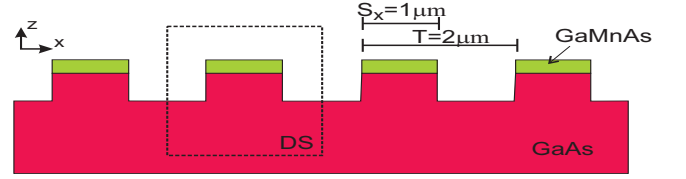}}%2.8cm
\caption{\label{fig(Ga,Mn)As_wire} (Color online) Schematic
cross-section of the GaMnAs/GaAs periodical wires perpendicular to
their direction (in the plane of the diffraction experiment).
The direct space (DS) area attained by the periodical object
diffracted signal sampling in RS is marked by the dashed line
region.}
\end{figure}

%Sample

Strain analysis in III-V semiconductor nanostructures composed of GaAs and (Ga,Mn)As is performed to demonstrate the potential of CXDI for invetigation of application-relevant nanoobjects.
Diluted magnetic semiconductors, of which (Ga,Mn)As is a prominent
example, have acquired great attention for tunability of their
magnetic properties and thus for their potential in spintronic
applications \cite{ZuticRMP2004, JungwirthRMP2006}. It has been
shown that, e.g., their magnetocrystalline anisotropy can be
substantially affected by elastic strain \cite{Wunderlich07}. One
way to induce controlled spatial variation of the strain is
lithographic patterning of the GaMnAs layer after the epitaxial
growth. Due to the positive lattice mismatch between (Ga,Mn)As and
GaAs, depending mainly on the concentration of Mn interstitial
atoms \cite{Masek03}, epitaxial (Ga,Mn)As layers deposited on GaAs
substrates are biaxially compressed. After creation of a periodic
sequence of thin wires and groves (surface grating), the film can
relax freely at the edges of the wires.

In the investigated structures the period of wires is $T=2$~$\mu$m
and the nominal wire width is $S_x=1$~$\mu$m. A 200~nm thick
(Ga,Mn)As layer with a nominal Mn concentration of about 7\% was
grown onto a (001)GaAs substrate. The periodic wire structure was
prepared by electron-beam lithography and reactive ion etching
with an etching depth of approx. 700~nm (see
Fig.~\ref{fig(Ga,Mn)As_wire}). The (Ga,Mn)As surface wires were
oriented along the $[1\overline{1}0]$ direction. The technological
details can be found elsewhere \cite{Wunderlich07}.

The wires do not contain large structure defects like
dislocations, stacking faults, or precipitates that would give
rise to considerable x-ray diffuse scattering. X-ray diffuse
scattering from point-like defects such as As$_{\rm Ga}$
antisites, Mn substitutional and interstitial atoms \cite{Masek03}
lies far below the detection limit of the experimental set-up.
Therefore, from the view-point of the experimental method used,
the wires represent an ideal crystal structure with a
two-dimensional distribution of the elastic displacement vector
${\bm u}(x,z)$ in the plane $xz$ perpendicular to the wire
direction.

%Experiment

Diffraction measurements were performed at ID10B beamline at
the ESRF. The x-rays from an in-vacuum undulator source were
monochromatized by a diamond(111) crystal to $E=8$~keV. The
cross-section of the primary beam was $0.2\times 0.2$~mm$^2$. High
resolution reciprocal-space images (RSI) were measured by use of a
Si(111) crystal analyzer. The result in vicinity of the 004 RLP is
shown in Fig.~\ref{figRSM004}. The RSI $q_xq_z$ plane is
perpendicular to the alignment of the wires.

Two well separated features in the 004 RSI correspond to the major
influence areas of the 004 reciprocal lattice points of GaAs and
(Ga,Mn)As respectively (the corresponding difference in Bragg
angles is about 0.2~degree). Grating maxima exhibiting a
horizontal spacing of about $\sim 3*10^{-4}$\AA$^{-1}$ can be
clearly resolved proving  the in-plane coherence length of the
incoming x-ray beam to be much larger than the 2~$\mu$m wire
period (see the magnified inset in Fig.~\ref{figRSM004}). Since
the lower sampling limit of RSM envelope corresponds to the rod
periodicity, the oversampling ratio of the measured signal in
$q_x$ direction is defined as $\sigma_x = \frac{T}{S_x}$. In our
case the wire width $S_x$ occupies the half of the grating period
$T$ and consequently the oversampling rate is about $\sigma_x = 2$
in $q_x$ direction. The internal symmetry of the wire in lateral
direction decreases the unknown information twice.

The dynamical diffraction peak of the semi-infinite GaAs substrate
cannot be taken into account since this region violates the
presumption of the phase retrieval method. The eliminated
experimental data near the GaAs reciprocal lattice point (marked
by $\Omega_{peak}$ in Fig.~\ref{figRSM004}) allowed to vary freely during reconstruction process~\cite{Nishino03}.

The support in direct space is chosen on the basis of growth
parameters and is finalized using Parseval theorem being applied
separately to the scattering regions $\Omega_L$ and
$\Omega_S$ in RSI shown in Fig.~\ref{figRSM004}.
Performing the phase-retrieval algorithm the Fouier components of electron
densities $\rho_{\bm h}({\bm r})$ of (Ga,Mn)As and GaAs are
assumed to be not affected by elastic deformation and shall vary
only with composition.

\begin{figure}                %Fig.
\resizebox{8.5cm}{8 cm}{\includegraphics{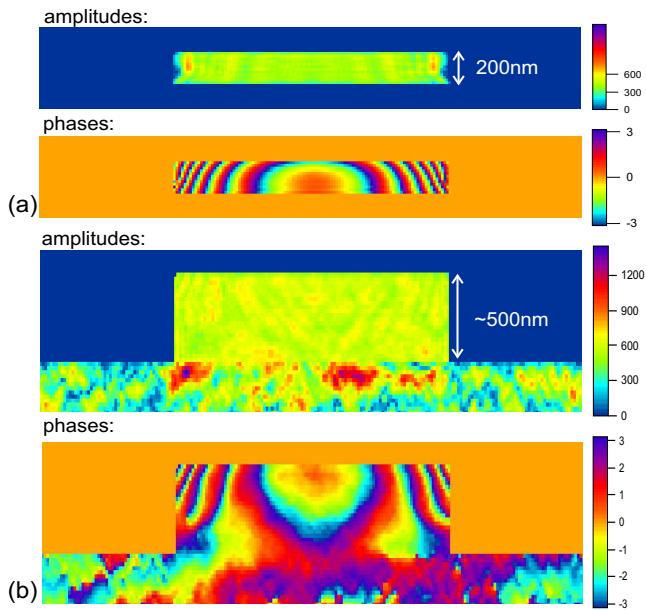}}%
\caption{\label{fig_reconstructions} (Color online) Results of
reconstructions (a) of $\Omega_L$ area of the RSI (see
Fig.~\ref{figRSM004}) corresponding to the (Ga,Mn)As part, (b) of
$\Omega_S$ area corresponding to the GaAs part. For the dimensions
of direct space of periodic objects see Fig.~\ref{fig(Ga,Mn)As_wire}.}
\end{figure}

From the diffraction pattern one can conclude on a large vertical
misfit between the (Ga,Mn)As and GaAs regions, allowing us to
divide the measured RSI in two regions $\Omega_L$ and $\Omega_S$
(see Fig.~\ref{figRSM004}). Each of them serves further for
independent reconstruction of phase and amplitudes in the
corresponding composite parts (Ga,Mn)As and GaAs of the wire,
respectively. The results are depicted in
Fig.~\ref{fig_reconstructions}. The reproducibility of the
obtained solutions was proven performing a series of inversion
cycles starting from different sets of random phases of the
scattered radiation. The error metric is defined as $E^2_k =
{\sum^{N}_{i=1} ( |F^{calc}_i| - \sqrt{I^{meas}_i}
)^2}/{\sum^{N}_{i=1} I^{meas}_i}$, where $|F^{calc}_i|$ is the
magnitude of the calculated amplitude and $I^{meas}_i$ is the
measured intensity of point $i$ in the RSI. The error metric is
found to be on the level of $10^{-3}$ for the (Ga,Mn)As part and
becomes even smaller, around $10^{-6}$, when inverting the GaAs
part. The direct-space resolution achieved is about $10\times
10$~nm$^2$.

\begin{figure}                %Fig.
\resizebox{8.5cm}{5 cm}{\includegraphics{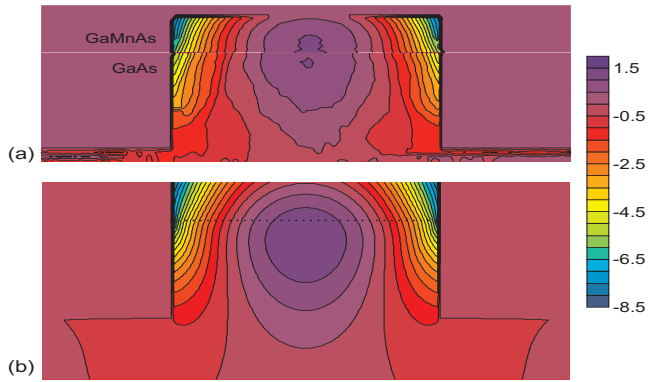}}%7cm
\caption{\label{fig_displacements} (Color online) Contour maps of
the vertical component $u_z$ of the displacement vector in the
(Ga,Mn)As/GaAs wire relative to the wire state before the
relaxation (in \AA) (a) directly reconstructed (calculated from
the two phases maps shown in Fig.~\ref{fig_reconstructions}), (b)
obtained numerically by finite-element simulations.}
\end{figure}

% results

From the results of the independent phase reconstructions of the
(Ga,Mn)As and GaAs volumes of the wires the following observations
can be made:

(i) During the reconstruction of the GaAs part the average
amplitude in the free-standing wire appears to be close to that one
in the (Ga,Mn)As (see Fig.~\ref{fig_reconstructions}).
This is in good agreement with the fact that the amplitudes of
structure factors of GaAs and (Ga,Mn)As unit cells containing
7\%~Mn differ by less than 1 percent.

(ii) The variation of the reconstructed phase fields (i.e., the
fields of the relative vertical displacement component
Fig.~\ref{fig_displacements}(a)) behave continuous at the boundary
between the (Ga,Mn)As and GaAs volumes. That fact further supports
the validity of our reconstruction process.

(iii) Finally, the combination of both independent reconstructions
into one leads after taking the FT to the same RSI as in the
Fig.~\ref{figRSM004} down to the error of $3*10^{-2}$. This shows
that even overlapping regions in RSI can be independently
reconstructed, when the difference in intensity levels between the
intensity maxima and the edge of the cut map is at least 2~orders
of magnitudes.

We do not interpret the non-homogeneous amplitudes in the GaAs
substrate part (under the wire). For this region strain variation
and compositional variation are certainly very small. Therefore
the missing signal data in the RSI close to the strong dynamical
peak (excluded from the diffraction pattern to be
reconstructed) is an essential loss of information and leads to
expected instabilities in the reconstructed amplitudes
in Fig.~\ref{fig_reconstructions}(b).

%results of FEM + details

Numerical finite-element method (FEM) calulations were performed
to simulate the strain field inside the (Ga,Mn)As/GaAs wires. Simulation used the
experimentally verified value of 0.35\% for the lattice mismatch
between the GaAs substrate and the (Ga,Mn)As layer containing 7\%
Mn.
%5.675-5.655=0.35%%
The Poisson ratio and the Young modulus of GaAs were used as an
approximation for (Ga,Mn)As. The results of the
simulations are shown in Fig.~\ref{fig_displacements}(b). The
excellent agreement between simulated (FEM) and
reconstructed strain field distributions proves the high quality
of the reconstruction process.

%Conclusion

In summary, we illustrated the ability of CXDI based on our
phase-retrieval algorithm to study non-uniform strain relaxation
in strained layer nanoobjects and for the first time succeeded to
reconstruct the related displacement fields.
Reproducible convergence of the CXDI method has been obtained even
for strongly inhomogeneously strained objects. The reconstruction
procedure takes benefit from the material-specific nature of CXDI
arising from compositionally driven lattice misfit. That enables
individual analysis of the corresponding compositional
constituents of the structure. The continuous field of
displacement can be also recovered by building up the strained
inhomogeneous system as a whole. The quantitative values of
displacement field reconstruction have been validated using the
elasticity theory. Sub ten nanometer resolution of the
displacement field in direct space was achieved. The
availability of a robust direct reconstruction method will allow
to gain precise knowledge about multicompositional strained
nanoobjects.

The work has been supported by
%the program PNI of the Helmholtz the Association, by
the NAMASTE project funded by the European Union.

%and by the Ministry of Education of the Czech Republic (program
%MSM 0021620834).
Authors thank to A. Singh and O. Konovalov for
preparing the experimental setup.

%The measurements were performed at ESRF. (has been said already in the text.

%\bibliography{biblio_CXDI}% Produces the bibliography via BibTeX.

\end{document}